\documentstyle[references, 10pt,epsf]{article}
\def\sze{SZE}
\def\aap{A\&A}
\def\apjl{ApJ}

\begin{document}

\centerline{\LARGE \bf{Imaging the Sunyaev-Zel'dovich Effect}}
\medskip
\centerline{
J. E. Carlstrom\footnote{\small University of Chicago, Department of Astronomy and Astrophysics,
5640 S. Ellis Avenue, Chicago, IL, 60637}
M. K. Joy\footnote{NASA/MSFC, Huntsville, AL, 35812},
L. Grego\footnote{Harvard-Smithsonian Center for Astrophysics, Cambridge, MA, 02138},
G. P. Holder$^1$}
\centerline{W. L. Holzapfel\footnote{University of California, Department of Physics, Berkeley, CA, 9472},
J. J. Mohr$^1$,
S. Patel\footnote{University of Alabama, Department of Physics, Huntesville, AL, 35899},
and E. D. Reese$^1$}

\medskip
\centerline{\small \it{to appear in the proceedings of the Nobel Symposium
``Particle Physics and the Universe"}}
\centerline{{\small \it Physica Scripta and World Scientific, eds.
L. Bergstrom, P. Carlson and C. Fransson}}

\begin{abstract}

We report on results of interferometric imaging of the
Sunyaev-Zel'dovich Effect (SZE) with the OVRO and BIMA
mm-arrays. Using low-noise cm-wave receivers on the arrays, we have
obtained high quality images for 27 distant galaxy clusters. We review
the use of the SZE as a cosmological tool.  Gas mass fractions derived
from the SZE data are given for 18 of the clusters, as well as the
implied constraint on the matter density of the universe, $\Omega_M$.
We find $\Omega_M h_{100} \le 0.22 ^{+0.05}_{-0.03}$.
A best guess for the matter density obtained by assuming a 
reasonable value for the Hubble constant and also by attempting to account
for the baryons contained in the galaxies as well
as those lost during the cluster formation process
gives $\Omega_M \sim 0.25$. We also give 
preliminary results for the Hubble constant.
Lastly, the power for investigating the high redshift
universe with a non-targeted high sensitivity SZE survey is discussed
and an interferometric survey is proposed.

\end{abstract}

\section{The Sunyaev-Zel'dovich Effect} 

The scattering of Cosmic Microwave Background (CMB) photons
by a hot thermal distribution of electrons leads to
a unique distortion of the CMB spectrum known as
the Sunyaev-Zel'dovich Effect (SZE) after the 
two Russian scientists who proposed it
in the early 1970's (\cite{sunyaev70}; \cite{sunyaev72}). 
In most cases,
and in all cases considered here, the hot gas
is provided by the intracluster medium of galaxy
clusters. For the most massive clusters the mass of
the intracluster medium greatly exceeds the mass contained
in the individual galaxies, although, as discussed in detail
below, roughly 85\% of the mass of a cluster is contained
in some other form of non-luminous matter. 
Even for the largest clusters with $\sim 10^{15}M_\odot$ total mass
the chance of a scattering a CMB photon traversing the cluster
is only about 1\%. This means that the resulting spectral
distortion will have a small amplitude.

\begin{figure}[t]
\epsfxsize= 5.0 in
\centerline{\epsfbox{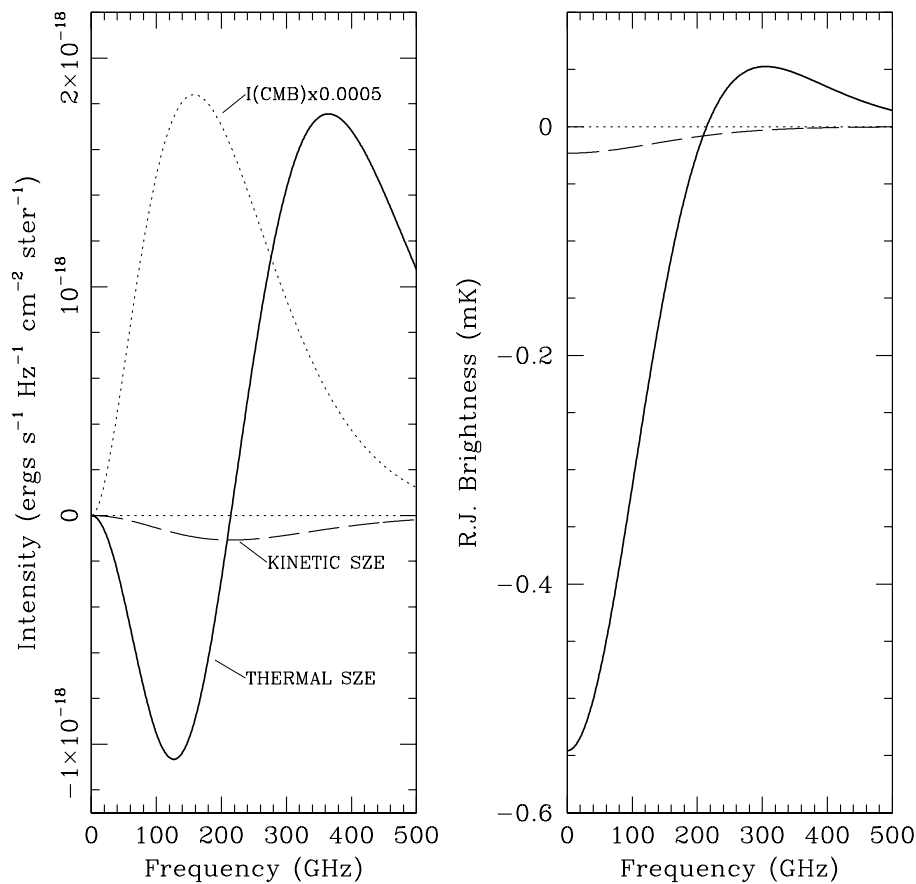}}
\caption{Spectral distortion of the Cosmic Microwave Background (CMB)
radiation due to the Sunyaev-Zel'dovich Effect (SZE). The left panel 
shows the intensity and the right panel shows the Rayleigh Jeans 
brightness temperature. The thick solid line is the thermal SZE and
the dashed line is the kinetic SZE. For reference
the 2.7~K thermal spectrum for the CMB intensity scaled by 0.0005 is shown by the dotted line
in the left panel; a line at zero is shown in both panels. The 
cluster properties used to calculate the 
spectra are an electron temperature of 10~keV, a Compton
$y$~parameter of $10^{-4}$ and a peculiar velocity of 500 $km\ s^{-1}$.}
\label{fig:spectrum}
\end{figure}

The photons that scatter off the much higher energy electrons
($T_{e} \sim 10^8$~K, 10 keV), will on average be shifted to
higher energy. The emergent spectrum is therefore distinctly
non-Planckian; compared to an initial Planck spectrum there are
fewer photons at low energies and more at high energies. 
The spectral distortion is
shown in Fig.~\ref{fig:spectrum} where the left panel shows the change
in intensity and the right panel shows the change in 
Rayleigh Jeans (RJ) brightness temperature. The RJ 
brightness is shown because the sensitivity of radio
telescope is calibrated in these units. It is defined simply
by $I_\nu = (2 k \nu^2/c^2) T_{RJ} \Omega$ where
$I_\nu$ is the intensity at frequency $\nu$, $k$ is Boltzmann's constant, 
$\Omega$ is the 
solid angle and $c$ is the speed of light. The Planck spectrum
of the CMB radiation is also shown by the dotted line
in Fig.~\ref{fig:spectrum} for reference. 

Also shown in Fig.~\ref{fig:spectrum} by the dashed curve 
is the kinetic \sze. This effect is caused by a non-zero
bulk velocity of the cluster with respect to the
CMB and along the line of sight; 
a peculiar velocity with respect to the Hubble flow. It
results in a purely thermal distortion of the
CMB spectrum (i.e., the emergent spectrum is still described completely
by a Planck spectrum, but at a slightly different
temperature, lower (higher) for positive (negative) peculiar velocities. 

The derivation of the exact spectral dependence of \sze\ can be found
in the original papers of Sunyaev and Zel'dovich as well as in a
number of more recent papers, which include relativistic corrections
to the earlier work, and reviews (e.g., \cite{sunyaev70}; \cite{sunyaev72};
\cite{sunyaev80}; \cite{rephaeli95}; \cite{birkinshaw99}).

At long wavelengths, the \sze\ toward massive and luminous galaxy
clusters should be observed as a hole in the sky relative to
the undistorted CMB brightness. This in itself is a remarkable
feature. There is no other plausible explanation known for
the existence of such a hole other than primary CMB anisotropy
which should have much smaller amplitudes at
the angular scales subtended by galaxy clusters
(e.g., \cite{holder99a}). A clear detection of a \sze\ decrement
was proposed as proof that the CMB was indeed cosmic (\cite{sunyaev70});
if proof is still needed we now know that it at least comes to
us from beyond $z = 0.83$.

In the Rayleigh-Jeans (RJ) limit the \sze\ spectral distortion is
given by
\begin{equation}
{\Delta T \over T_{CMB}} = -2\int {k T_e \over m_e c^2} \sigma_T n_e dl
\label{eq:spectrumRJ}
\end{equation}
where $T_{CMB}$ is the radiation temperature of the CMB, $k$ is Boltzmann's 
constant, $n_e$ and $T_e$ are the electron density and 
temperature, respectively, $\sigma_T$ is the Thompson cross section,
$m_e$ is the mass of the electron, $c$ is the speed of light
and the integral is along the line of sight.
Using the definition
for the Comptonization parameter $y$, we find that $ {\Delta T \over T_{CMB}} 
= -2 y$ in the RJ limit.

Perhaps the most amazing fact of the \sze\ is best illustrated by
Eq.~\ref{eq:spectrumRJ}; the observed brightness of the effect $\Delta T$
is independent of the distance (redshift) to the cluster! 
Both $\Delta T$ and  $T_{CMB}$ suffer the same cosmological
dimming, so their ratio is simply a function of the cluster properties.
And, of course, we observe the same $T_{CMB}$ toward
any cluster at any redshift. The integrated \sze\ flux
from a cluster depends on $\Delta T D_A^{-2}$ and is thus dependent
on distance. For observations in which the angular resolution
is sufficient to resolve the effect, however, the observable is independent
of redshift. This requirement is obtained easily, as clusters
are large objects ($\sim$Mpc) and therefore subtend 
an arcminute or more at any redshift, assuming a reasonable cosmology.
This wonderful property of the \sze\ makes it a potentially 
powerful probe of the high redshift universe.

As discussed in section~\ref{sec:cosmo}, sensitive observations of 
the \sze\ provide a powerful and unique cosmological tool. 
Our group has used interferometric techniques to make high
quality images of the \sze\ toward more than 25 clusters
with redshifts spanning 0.13 to 0.83. In section~\ref{sec:obs},
we give a brief review of our observing technique and present
some of the resulting images. We discuss our progress
on the Hubble constant and mass density of the universe in
section~\ref{sec:results}, and finally in section~\ref{sec:conclusions}
we briefly review our future plans.

\section{Cosmology with the Sunyaev-Zel'dovich Effect}
\label{sec:cosmo}

Beyond the novelty of measuring a hole in the sky, or adding to the
already overwhelming evidence that the CMB is indeed cosmic,
measurements of the \sze\ offer a powerful and unique way to test
cosmological models and determine the values of the cosmological
parameters which describe our universe.

Here we concentrate on the \sze\ from galaxy clusters, 
the largest known collapsed objects in the universe. Galaxy clusters
themselves provide important sign-posts of structure in the universe.
Their properties and evolution provide valuable constraints
on cosmological models. In addition to its \sze,
the hot intracluster gas is traced by its strong bremsstrahlung radiation
at X-ray wavelengths. The deep gravitational potential can be probed
by X-ray spectroscopy (to measure of the intracluster 
gas temperature), by the velocity dispersion
of member galaxies, and by the gravitational lensing of
background galaxies. 

We now discuss briefly the cosmological probes that are offered
directly by \sze\ measurements of clusters, both by itself and when
used in conjunction with the measurements discussed above.

\subsection{Hubble constant; expansion history of the universe}

Perhaps the \sze\ is best known for providing a means to measure the
distance to a galaxy cluster independent of any other distance scale;
it does not require normalization to distances derived for more nearby objects 
as is the case with the ``distance ladder'' that is commonly used.
The distance is derived by combining a measurement of
the \sze\ with a measurement of the X-ray intensity of the cluster. To
understand how this is possible it is only necessary to consider the
different dependencies of the \sze\ and X-ray observables on the
electron density of the intracluster gas.  

The \sze\ is simply
dependent on the integrated density as indicated in
Eq.~\ref{eq:spectrumRJ}.  The X-ray intensity is proportional to the
density squared as given by
\begin{equation}
I_X(E,\delta_E) = {1 \over 4 \pi (1+z)^4} {\mu_e\over \mu_H} \int n_e^2 \Lambda(E',\delta_{E}',T_e) dl
\label{eq:X-ray}
\end{equation}
where $I_X(E,\delta_E)$ is the X-ray intensity observed 
within a fixed detector bandwidth $\delta_E$ at energy $E$,
$\mu_j \equiv \rho/(n_j m_p)$, $\rho$ is the gas mass density,
$m_p$ is the mass of the proton, $\Lambda(E',\delta_E',T_e)$ is
the emissivity within a bandwidth $\delta_{E}'$ at energy $E'$
of a $T_e$ gas 
in the
cluster rest frame ($E' = (1+z)E,\ \delta_{E}' = (1+z)\delta_{E}$), and the integral is again along
the line of sight.
Note, 
$\Lambda(E',T_e)$ decreases steeply with $z$ over the energy
range of interest (usually in the range 0.5 to 10~keV), 
so that the detectability
of a cluster for a given X-ray detector typically falls steeper than
$1/(1+z)^4$ in sharp contrast to the essentially
redshift independence of the \sze\ signal.
 
Due to the different electron density dependencies of the observed
\sze\ ($\propto n_e$) and X-ray ($\propto n_e^2$) emission, one can use the measurements to
constrain the electron distribution, or at least constrain the
parameters within a given model for the gas. One also needs 
the electron temperature which can be measured using X-ray spectroscopy.
The determined gas distribution can then be compared with the
measured angular distribution to solve for the distance to
the cluster. A comparison with the mean redshift of the 
member galaxies gives the Hubble constant $H_o$. 

The beauty of this technique for measuring the Hubble constant is
that it is completely independent of other techniques, and
that it can be used to measure distances at high redshifts. While the method depends
only on well understood properties of fully ionized plasmas, there
are several sources of uncertainty in the derivation of
the Hubble constant for any particular cluster (e.g., 
see \cite{birkinshaw99}). The largest uncertainty
is that we are making the assumption that the cluster size
along the line of sight is comparable to its size
in the plane of the sky. For this reason it is desirable to 
use a large survey of clusters to determine $H_o$. A large survey
of perhaps a few hundred clusters with redshifts ranging from
close by to beyond one would allow the technique to be used to
trace the expansion history of the universe, providing
a valuable independent
check of the type Ia supernova results (e.g., \cite{riess98}; \cite{perlmutter99}).

\subsection{Peculiar velocities}

The line of sight velocity of a cluster with respect to
the CMB rest frame, the peculiar velocity of the cluster,
can be measured by separating the kinetic from the
thermal \sze. From inspection of Fig.~\ref{fig:spectrum}, it
is clear that this is best done by observation at frequencies
near the null of the thermal effect at $\sim 218$~GHz. Such measurements
offer the ability to measure the peculiar velocity of clusters
at high redshifts which could be used to constrain the 
large scale gravitational perturbations to the Hubble flow.
The intrinsic weakness
of the effect make the observation of the effect challenging. 
Upper limits have been
placed on the peculiar velocities of clusters (\cite{holzapfel97b}),
but a clear detection of the kinetic effect has not yet been
obtained. The kinetic \sze\ is a unique and potentially
powerful cosmological tool as it provides the only known way
to measure large scale velocity fields at
high redshift. We are likely to see continued progress in
these difficult observations. Our \sze\ observations
discussed later in this paper were made at 30 GHz where
the kinetic effect is clearly of second order to the thermal effect,
and it is therefore not discussed further here. 

\subsection{Baryon mass fraction of clusters; $\Omega_M$}

A measurement of the \sze\ toward a cluster provides a measure of the
mass of the intracluster medium, which is typically several times the
mass responsible for the light from the galaxies. Combining the gas
mass with a measure of the total mass determined either from
gravitational lensing observations or from the virial theorem
and the X-ray determined electron temperature, one can
determine the fraction of the mass of the galaxy cluster contained in
baryons. An estimate of the baryonic to total mass on the scale of
massive galaxy clusters is important as it should represent the
universal value; it is not believed that mass segregation occurres on
the scales from which massive clusters condense ${\rm \sim 1000\ Mpc^3}$,
although as noted below a small fraction of baryons ($\sim$15\%) are likely 
lost during the cluster formation process.

The universal mass fraction of baryons to total matter,
$\Omega_B/\Omega_M$ where $\Omega \equiv \rho/\rho_c$, and $\rho_c$ is
the critical density of the universe, can in turn be used to estimate
$\Omega_M$ given the value determined for $\Omega_B$ from big bang
nucleosynthesis calculations and the observed values for the
primordial abundance of the light elements (\cite{burles98a}).

\subsection{Cluster evolution; probing the high redshift universe}

\begin{figure}[t]
\epsfysize = 3.0 in
\centerline{\epsfbox{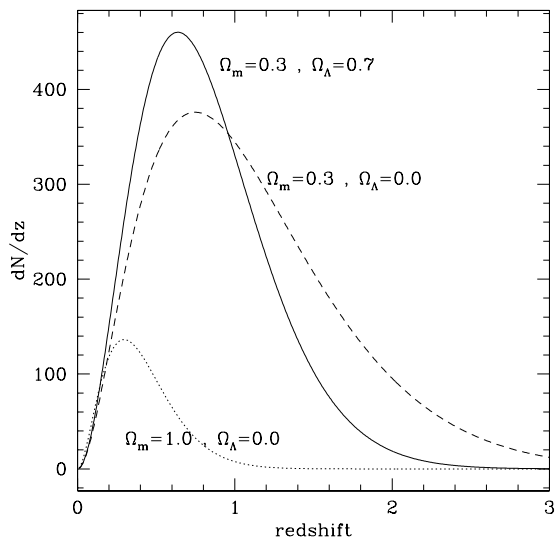}}
\caption{The predicted $z$ distribution of clusters
with masses greater 
${\rm 2\times 10^{14}\ M_\odot}$
for several cosmologies (Holder and Carlstrom 1999). The numbers 
are for a 12 square degree patch of sky. 
Note the cosmological sensitivity of the cluster redshift distribution.}
\label{fig:dndz}
\end{figure}

Perhaps the most powerful use of the \sze\ for cosmology will
be to probe the high redshift universe. 
Sensitive, non-targeted surveys
of large regions of the sky for the \sze\ will provide an 
inventory of clusters independent of redshift. The 
cut-off of the \sze\ cluster sample would be a lower mass cut-off 
set by the sensitivity
of the observations; there would be no redshift cut-off. 
The sensitivity of a such a survey to cosmological
parameters is shown in Fig.~\ref{fig:dndz}, where the predicted
distribution in redshift of all clusters with 
masses greater than $2\times 10^{14}\ M_\odot$ is shown
(\cite{holder99b}).
A sufficiently
sensitive \sze\ survey would also be able to image the ionized gas
expected in filamentary large scale structure, particularly the filaments
associated with the formation of clusters. 

The use of the number density of clusters, particularly massive
clusters, as a function of redshift
has been used by a number of authors to estimate $\Omega_M$
from X-ray surveys (e.g., \cite{bahcall99}). The possible use of \sze\ surveys to constrain 
$\Omega_M$ has been investigated as well (\cite{barbosa96}; \cite{colafrancesco97}) along with more recent studies which include the effect of
the cluster gas evolution (\cite{holder99a}).

\section{\sze\ Observations}

\label{sec:obs}

\subsection{Previous observations}

In the twenty years following the first papers by
Sunyaev and Zel'dovich, there were few firm detections
of the \sze\ despite a considerable amount of effort
(\cite{birkinshaw91}). Over
the last several years, however, observations of the
effect have progressed from low
S/N detections and upper limits to high confidence detections and
detailed images. The dramatic increase in the quality of the
observations is due to improvements both in low-noise detection
systems and in observing techniques, usually using specialized
instrumentation with which the systematics that often prevent one
from obtaining the required sensitivity are carefully controlled. Such
systematics include, for example, the spatial and temporal variations
in the emission from the atmosphere and the surrounding ground.

A recent review of the observations can be found in 
\cite{birkinshaw99}. Here we briefly review a few of the
results to provide the reader with a measure of the
quality of the data presently available. We then concentrate
on our own technique and imaging results.

The first measurements of
the \sze\ were made with single dish radio telescopes. 
Successful detections were obtained, although the reported
results show considerable scatter, reflecting the difficulty of the
measurement. Recent state-of-the-art single dish
observations at radio wavelengths (\cite{herbig95}; \cite{myers97}),
millimeter wavelengths
(\cite{wilbanks94}; \cite{holzapfel97a}) and submillimeter
wavelengths (\cite{lamarre98})
have resulted in significant detections of the effect
and limited mapping. 

Interferometric techniques have been used to produce high quality
images of the \sze\ (e.g., \cite{jones93}; \cite{grainge93};
\cite{carlstrom96}; \cite{grainge96}; \cite{carlstrom98}). 
As discussed in the 
next section, the high stability and spatial filtering made possible
with interferometry is being exploited to make these
observations.

\subsection{Interferometry basics}

The stability of interferometry is attractive for avoiding many of the
systematics which can prevent one from imaging very weak emission.
The `beam' of a two-element interferometer -- all arrays can be
thought of as a collection of $n(n-1)/2$ two-element interferometers
-- is essentially a cosine corrugation on the sky; it is exactly
analogous to a two slit interference pattern. The interferometer
does the job of multiplying the sky brightness at the observing
frequency by a cosine, integrating the
product and producing the time average amplitude. The correlator
performs the multiplication and time averaging. In practice two
correlators are used to obtain the cosine and sine patterns. Simply
put, the interferometer measures directly the Fourier transform of the
sky at a spatial frequency given by $B/\lambda$, where
$B$ is the component of the vector connecting the two telescopes
(the baseline) oriented perpendicular to the source. Of
course, a range of baselines are actually being used at any one time
due to the finite size of the apertures of the individual array
elements; this simply reflects that the sky has been multiplied by the
gain pattern (beam) of the individual telescopes or, equivalently,
that the Fourier transform measured is the transform of the true sky
brightness convolved with the transform of the beam of an array
element.

That the Fourier transform measured is the transform of the 
sky brightness convolved with the
transform of the beam of an individual element has important consequences.
The transformed beam is the auto-convolution of the aperture, and thus it
is identically zero beyond the diameter of the telescopes expressed
in wavelengths. The interferometer is therefore only sensitive to angular scales 
(spatial frequencies) near $B/\lambda$. It is not sensitive to
gradients in the atmospheric emission or other large scale emission
features. 
There are several other features which allow an interferometer to
achieve extremely low systematics. For example, only signals which correlate
between array elements will lead to detected signal. For most interferometers,
this means that the bulk of the sky noise for each element will not
lead to signal. Amplifier gain instabilities for an interferometer
will not lead to large offsets or false detections, although if
severe they may lead to somewhat noisy signal amplitude. To remove
the effects of offsets or drifts in the electronics as well
as the correlation of spurious (non-celestial) sources of noise,
the phase
of the signal received at each telescope is modulated 
before the correlator and then the 
proper demodulation is applied to the output of the
correlator. 

Lastly, the spatial filtering of an interferometer allows the emission
from radio point sources to be separated from the \sze\ emission. This
is possible because at high angular resolution ($<10''$) the \sze\
contributes very little flux.  This allows one to use long baselines
-- which give high angular resolution -- to detect and monitor the
flux of radio point sources while using short
baselines to measure the \sze.
Nearly simultaneous monitoring of the point sources is
important as they are often time variable.  The signal from the point
sources is then easily removed, if they are not too strong, from the
short baseline data which are sensitive to the \sze.  Nevertheless, one
would still prefer to operate at shorter radio wavelengths since the
point sources typically have falling spectra ($\sim \lambda^{0.7}$),
while the \sze\ signal scales as ($\lambda^{-2}$).

For the reasons given above, interferometers offer an ideal
way to achieve high brightness sensitivity for extended 
low-surface brightness sources, at least at radio wavelengths. 
Most interferometers, however, were not designed for
imaging low-surface brightness sources.
Interferometers
are traditionally built to obtain high angular resolution with
large individual
elements for maximum sensitivity to small scale emission. Galaxy clusters,
on the other hand are large objects.  Most of the \sze\
signal will be distributed smoothly on angular scales of
an arcminute or more for even the most distant clusters, scales for
which most existing interferometric arrays are simply not sensitive.

\subsection{OVRO and BIMA interferometric imaging of the \sze}

\begin{figure}
\epsfysize= 7.0 in
\centerline{\epsfbox{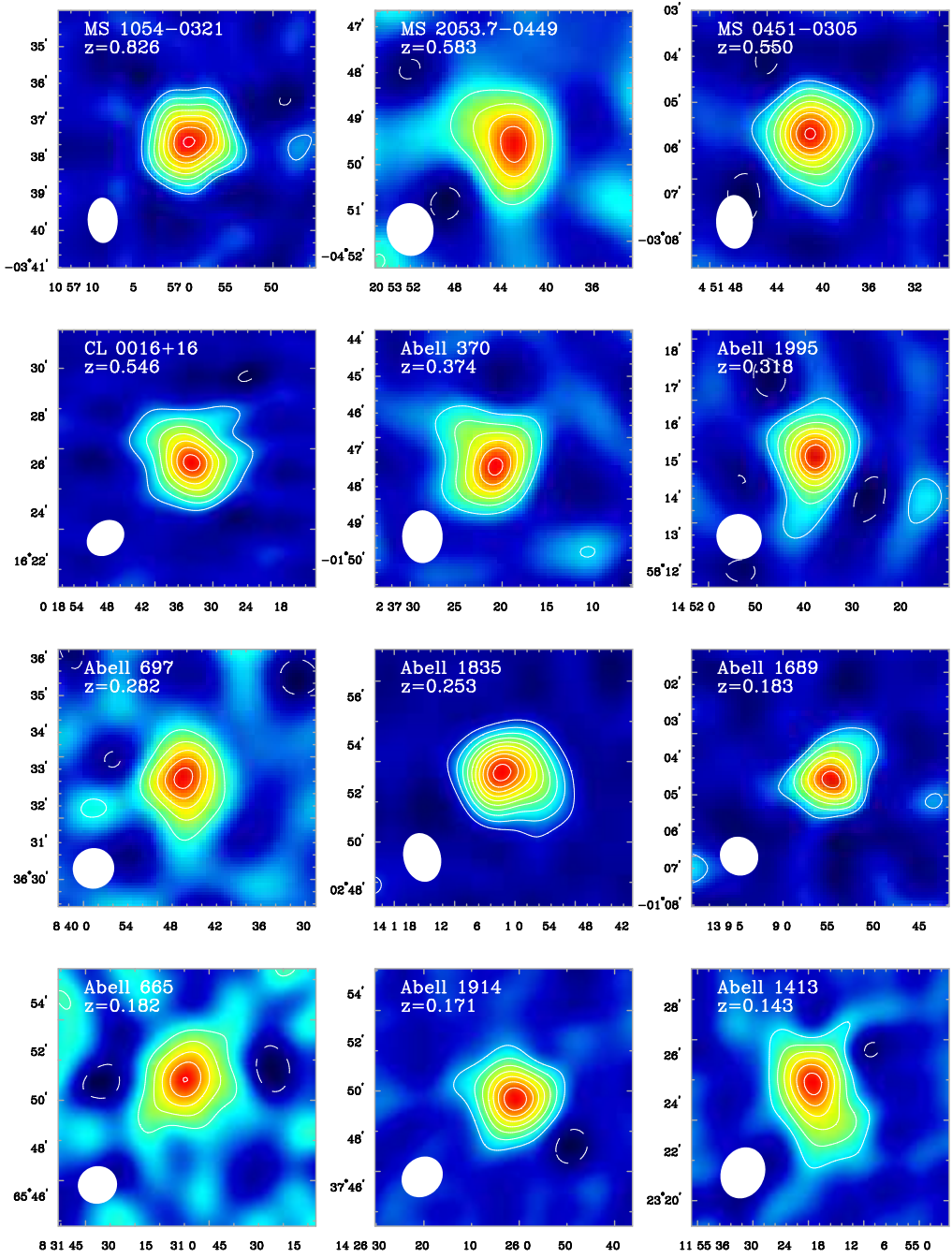}}
\caption{Images of the Sunyaev-Zel'dovich effect toward twelve distant
clusters with redshifts spanning 0.83 (top left) to 0.14 (bottom right). The evenly spaced
contours are multiples starting at $\pm 1$ of 1.5$\sigma$ to 3$\sigma$ depending
on the cluster, where
$\sigma$ is the rms noise
level in the images. The noise levels range from 15 to 40~$\mu$K.  The
data were taken with the OVRO and BIMA mm-arrays outfitted with low-noise
cm-wave receivers. The filled ellipse shown in the bottom left
corner of each panel represents the FWHM of the effective resolution
used to make these images.}
\label{fig:szpanel}
\end{figure}

\begin{figure}
\epsfysize = 3.5 in
\centerline{\epsfbox{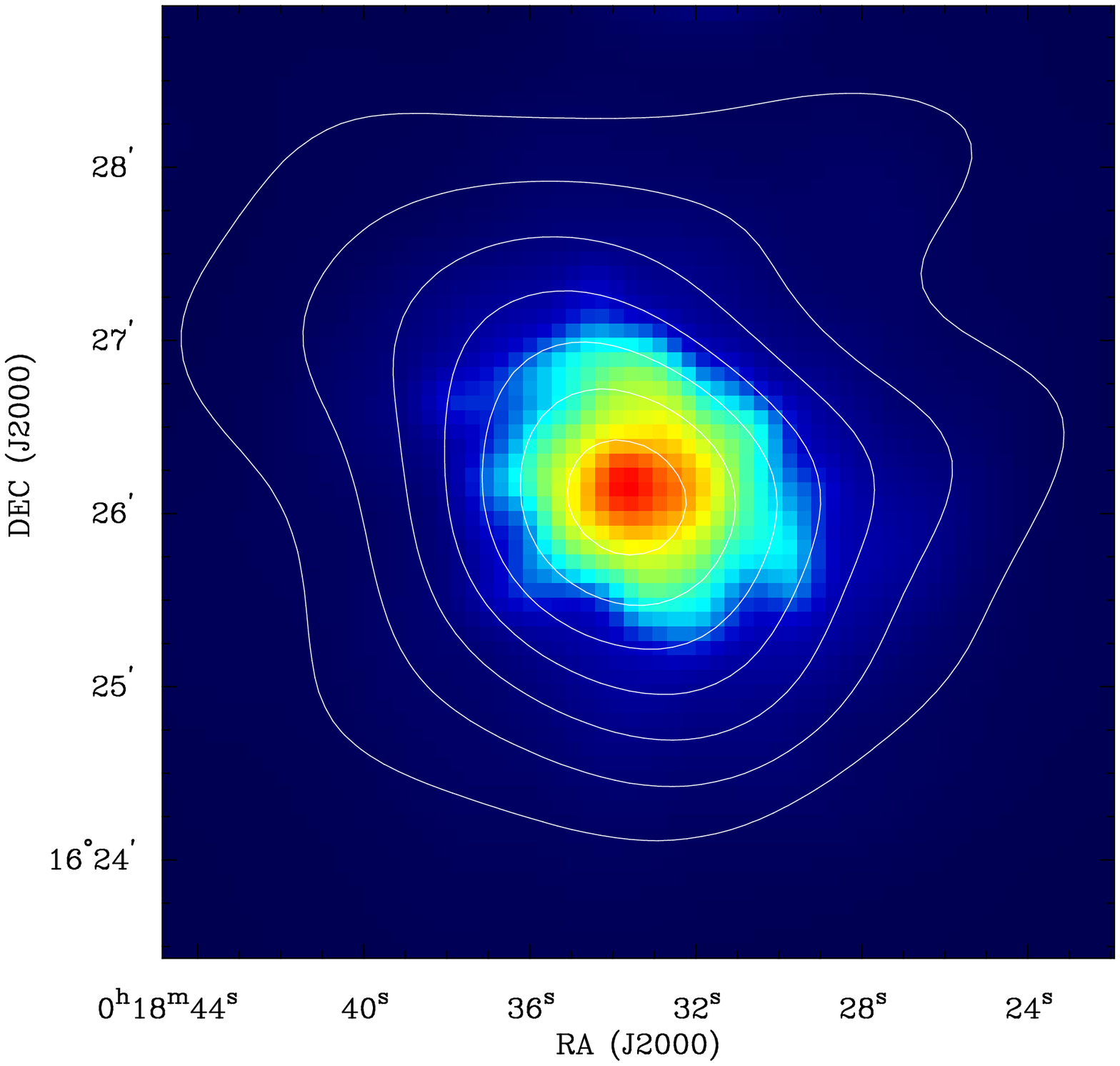}}
\caption{The Sunyaev-Zel'dovich effect (contours) overlaid
on the X-ray emission (false color) for the galaxy cluster
CL~0016+16. The \sze\ data were obtained with the 
OVRO and BIMA mm-arrays. The X-ray data were obtained with
the ROSAT PSPC.}
\label{fig:cl0016-xray}
\end{figure}

Our solution to matching the angular scales important for
observations of distant galaxy clusters (a few arcminutes) with
those provided by an interferometer is to use existing
mm-wave arrays, which were also designed for high resolution,
but to degrade the angular resolution by outfitting the 
arrays with cm-wave receivers. This solution has a number of
attractive features in addition to matching the arcminute
scales appropriate for \sze\ measurements of distant clusters.
Specifically, we are able to use very low noise cm-wave
receivers. We are able to secure large amounts of observing
time during the summer months when the atmosphere is not
ideal for mm-wave observations, but reasonable for cm-wave.
As one might imagine, there are also a number of advantages
to using telescopes which have surface and pointing
accuracies ten times higher than needed.

\setcounter{footnote}{0}

Over the last few summers we have thus installed low-noise, HEMT amplifier based
receivers on the OVRO\footnote{An array of six 10.4 m mm-wave telescopes
located in the Owens Valley, CA and operated by Caltech.} and BIMA\footnote{An array of
ten 6.1 m mm-wave telescopes located
at Hat Creek, California and operated by the 
Berkeley-Illinois-Maryland-Association} mm-wave arrays in California. The
receivers operate from 26 - 36~GHz (30 GHz corresponds to 1 cm), and
down convert the signal to the standard intermediate frequencies used
at the two arrays.  All of the normal array correlators, electronics,
and software are used. About 1~GHz can be processed at one time
with the standard observatory electronics.  Data are taken from array
configurations in which roughly half the baselines are sensitive only
to point sources and the other half are as short as possible for
maximum sensitivity to the \sze.

In Fig.~\ref{fig:szpanel} we show a subset of the 27 clusters
that we imaged so far using the OVRO and BIMA arrays. Contaminating
emission from radio point sources was removed before making these images.
The short baseline data were emphasized when these images
were made to enhance the surface brightness sensitivity. The
data do contain significant information at higher resolution.
A catalog of the point sources we measured toward our \sze\ imaged clusters 
can be found in \cite{cooray98}. Fig.~\ref{fig:szpanel}
illustrates the high quality data we have been able to obtain. The typical
integration time used for each cluster is about 45 hours, which
when one includes calibration and other overheads takes roughly
8 to 10 transits of the source.

Fig.~\ref{fig:szpanel} also clearly demonstrates the independence
of the \sze\ on redshift. All of the clusters shown have similar
high X-ray luminosities and, as can be seen, the strength of
the \sze\ is similar for each one. For the associated X-ray emission,
however,
the rate of received photons falls off drastically with redshift.

The \sze\ image is shown (contours) on the corresponding
X-ray emission in Fig.~\ref{fig:cl0016-xray}. As expected, the
\sze\ which traces the density of the cluster gas 
is less centrally peaked than the X-ray emission which traces the
square of the density.

\begin{figure}[t]
\epsfysize= 3.5 in
\centerline{\epsfbox{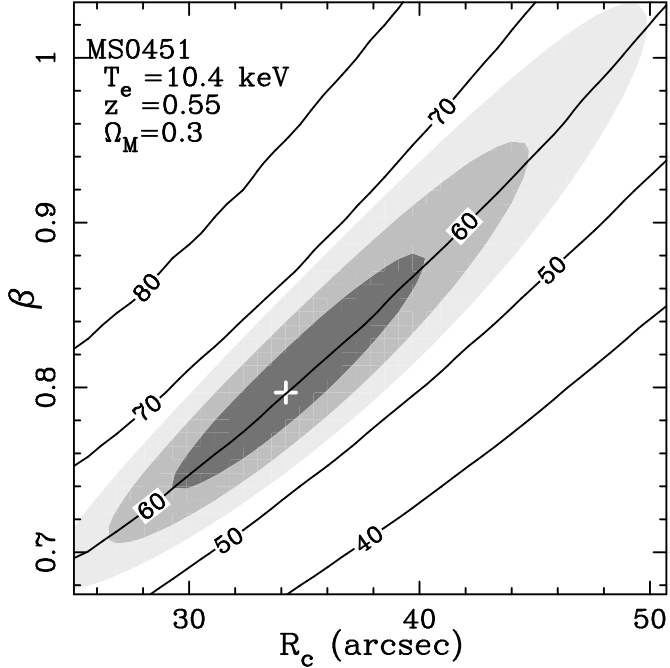}}
\caption{The derived Hubble constant in units of $km\ s^{-1} Mpc^{-1}$
is shown as contours from
joint fits to Sunyaev-Zel'dovich data and X-ray data to the
cluster MS~0451-0305. The 1,2 and 3$\sigma$ confidence
regions for the fit parameters $\beta$ and $R_c$ are 
shown in greyscale. Note that the value of the 
fitted Hubble constant is essentially insensitive to
the clear degeneracy between the fitted values of $\beta$ and $R_c$.
}
\label{fig:hubble}
\end{figure}

\section{Results}
\label{sec:results}

We have been developing methods for best extracting the 
cosmological parameters from our \sze\ data. One has to
keep in mind that the images shown in Fig.~\ref{fig:szpanel},
while they give a direct indication of the data quality, have
been heavily filtered by the response of the interferometer. 
Therefore, we do all of our analyses in the Fourier domain
where the data are actually measured. 

We fit models for
the cluster gas distribution to our data by first constructing
a realization of the model, projecting it in the plane
of the sky, multiplying it by the angular gain response of
our instrument (the beam of an array element), and then 
Fourier transforming it to the spatial frequency of our
data points. A comparison of the model Fourier transform
points and the data points is used to construct the likelihood
of the model.

For a suitable model we start with a standard $\beta$-model
for the gas distribution
\begin{equation}
n_e(r) = n_e(0)\left(1+{r^2\over R_c^2}\right)^{-3\beta/2}
\label{eq:beta}
\end{equation}
where $n_e(0)$ is the central electron density, $R_c$ is the
core radius, and $\beta$ a power law index. In practice, we
generalize this to be elliptical, fitting for the position
angle and major/minor axes ratio. We also fit the location
of the cluster as well as the locations and fluxes of any radio
point sources.  We assume the gas is isothermal.
For each set of parameters a likelihood (actually
a $\chi^2$ since the noise is Gaussian) is obtained. 

When X-ray data is also used in the analyses, a similar procedure
is followed and the joint likelihood is obtained. 

\begin{figure}[thb]
\epsfysize= 3.0 in
\centerline{\epsfbox{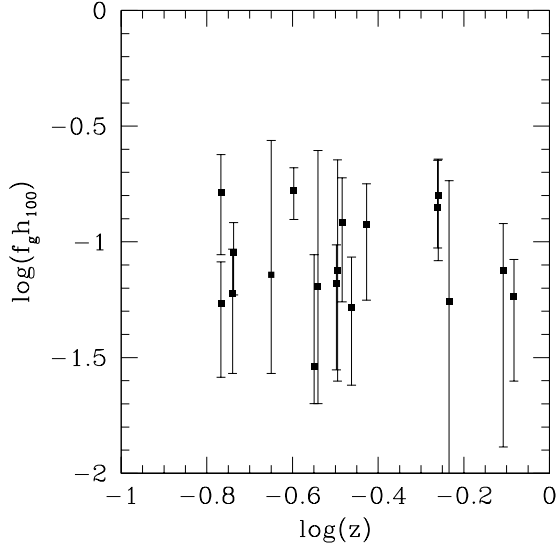}}
\caption{The gas mass fraction for galaxy clusters
derived from OVRO and BIMA Sunyaev-Zel'dovich effect data as a function of redshift.}
\label{fig:fg}
\end{figure}

\subsection{Hubble constant}

We are performing joint analyses of X-ray and our \sze\ data
to determine the distance to each cluster. The result of
such a derivation for the cluster MS~0451-0305 is 
shown in Fig.~\ref{fig:hubble} (\cite{reese99}). The values for $H_o$
(in units of $km\ s^{-1} Mpc^{-1}$) are 
shown as contours on a plot of the model parameters
$\beta$ and $R_c$. The 1, 2 and 3$\sigma$ confidence
intervals are shown by the greyscale regions. Note
the striking independence of the derived value of $H_o$ along
the clear degeneracy in the fits to these model parameters.

There are uncertainties in the absolute calibration of
both the \sze\ and X-ray data. However, the largest uncertainty
in the derived $H_o$ for any one cluster is the unknown
cluster aspect ratio, since we implicitly assume the cluster
dimension along the line of sight is equal to its dimension
across the line of sight. That our analyses
of MS~0451-0305 gives a value for $H_o$ that is so close
to the currently accepted range for $H_o$ is somewhat
fortuitous. To obtain a trustworthy estimate for
$H_o$ using the \sze\, 
one must obtain a large sample of distances
and be careful of selection effects. We are in the 
process of conducting such a survey now. In his recent
review, Birkinshaw found the average for 
all published \sze\ derived values of $H_o$ to be
$60 \pm 10\ km\ s^{-1} Mpc^{-1}$ 
(\cite{birkinshaw99}). Note, however, as Birkinshaw
points out, that the underlying
observations share common and fairly uncertain calibrations.

\subsection{Cluster gas mass fractions and constraints on $\Omega_M$}

We have determined the gas mass fractions for 18 clusters using only
our \sze\ data and electron temperatures derived from X-ray
spectroscopy (\cite{grego99a}; \cite{grego99c}). We did not use X-ray
imaging data, and in that sense our results provide a test of similar
analyses done on X-ray data sets.  Details of our technique in which
we account for the expected bias of the gas mass fraction in the core region of
a cluster can also be found in \cite{grego99b}.  The resulting gas mass
fractions as a function of redshift are shown in
Fig.~\ref{fig:fg}. In most of the clusters, the uncertainty is 
dominated by poorly
constrained electron temperatures.

The degeneracy in the fit parameters $R_c$ and $\beta$ is worse
than that shown in Fig.~\ref{fig:hubble} since we are not
using the X-ray data to help constrain the fits. We 
find that the derived gas mass and
the gas mass fraction are insensitive to the particular values of
$R_c$ and $\beta$ within the region
of acceptable fits, as long as these quantities are only calculated within 
about an arcminute radius region. This is not
surprising,  as the \sze\ signal is directly proportional to the mass 
and our data constrains the signal well on these scales.  We relate 
the gas fractions we derive at arcminute scales to the expected gas
fractions near the cluster virial radius using scaling relations
derived from numerical simulations of cluster formation
\cite{evrard97}).

\begin{figure}[thb]
\epsfysize= 3 in
\centerline{\epsfbox{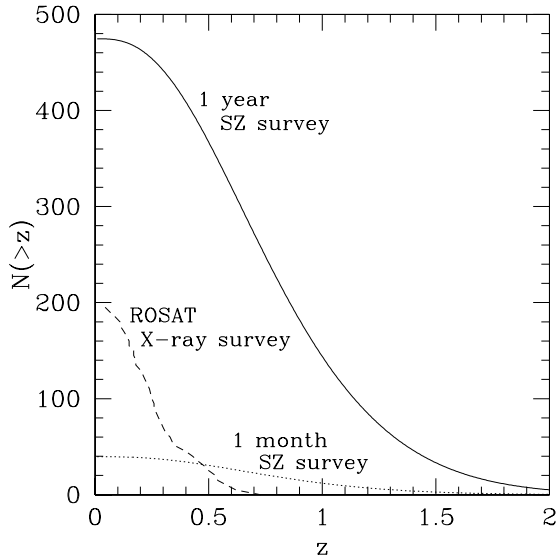}}
\caption{The expected number of clusters detected with an array of 10
2.5-m telescopes operating at 1~cm and with a correlation bandwidth of
8~GHz conducting a survey for one month-- 1 deg$^2$-- (dotted) and for
one year-- 10 deg$^2$-- (solid). The predictions are based on a 
$\Omega_\Lambda = 0.7$, $\Omega_M = 0.3$ flat CDM
cosmology; the predicted numbers are comparable for an open CDM model. The
resulting sample of clusters would have more than 300 clusters with $z>0.5$
and more than 100 with $z>1$. For comparison, we
show the cluster sample from a deep, serendipitous, X-ray survey of
$\sim$160~deg$^2$ carried out using archival ROSAT PSPC images.}
\label{fig:szcounts}
\end{figure}

The eighteen clusters range in redshift from 0.171 to 0.826.  Assuming
an open $\Omega_M = 0.3,\ \Omega_\Lambda = 0$ cosmology, we find $f_g h_{100} =
0.075 ^{+0.007}_{-0.010}$ scaled to $r_{500}$, the radius at which the
overdensity of the cluster is 500, nearly the virial radius. 
Using the six clusters with redshifts
less than 0.26, so that the cosmology we assume will have little effect
on the gas mass fractions derived, we find $f_g h_{100} = 0.085
^{+0.011}_{-0.015}$, again scaled to $r_{500}$.

Our results are in good agreement with those derived by \cite{myers97}
for a sample of 3 nearby clusters, and also with values derived from
X-ray emission (\cite{david95}; \cite{mohr99}), suggesting that
systematic
differences between the methods, like the clumping of intracluster
gas, are not severe.

Assuming that the baryonic mass fraction for clusters reflects the
universal value and using the constraint on $\Omega_B$ determined 
from BBN and 
primordial abundance measurements (\cite{burles98a}), $\Omega_B
h_{100}^2 = 0.019 \pm 0.001$, the low redshift gas mass fractions 
imply $\Omega_M h_{100} \le \Omega_B/f_g = 0.22 ^{+0.05}_{-0.03}$.
A best guess of $\Omega_M$ can be made by attempting to account
for the baryons lost during the cluster formation process (15\%) 
and for the baryons contained in the galaxies, as well
as using a best guess for the Hubble constant. This gives 
$\Omega_M \sim 0.23^{+0.06}_{-0.04} h^{-1}_{65}$. The uncertainties
should be taken with caution as they do not reflect the
uncertainties in the assumptions used to extrapolate from
$\Omega_B/f_g$ to $\Omega_M$.

\section{Discussion and Future Plans}
\label{sec:conclusions}

We are able to obtain high quality imaging data of the
\sze\ toward distant galaxy clusters. Our constraint
on the gas mass fractions of galaxy clusters provides further
support that we live in a low $\Omega_M$ universe. Our $H_o$
results are still preliminary, but show promise of providing
a good independent estimate for the expansion of the universe.

We are continuing to work on our analysis tools, including testing
thoroughly the effects of our assumptions such as the isothermality of
the cluster gas and the limits of $\beta$-models.

We are still a long way from being able to constrain whether
the universe is accelerating or not. And, even though our
sensitivity is quite high, we are not yet in a position to
conduct large sensitive, non-targeted surveys for the \sze\
over large regions of the sky. We have, however, clearly demonstrated
the feasibility of using similar interferometric techniques to conduct
such a survey.

A dedicated array with 10 elements of 2.5 m diameter
and using our cm-wave receivers would increase the speed of 
system dramatically. In Fig.~\ref{fig:szcounts} we show
the expected number of clusters detected with such
an instrument (see \cite{holder99b}). 
One month of 
observing with this array
would deliver more clusters with redshifts higher than $z \sim  0.5$
than are found in the deepest, large area X-ray cluster
catalogs (\cite{vikhlinin98}).  The plot assumes a
$\Omega_m=0.3$, $\Omega_{\Lambda}=0.7$ cosmological model; 
if $\Omega_{\Lambda}=0$ the counts are comparable (see Fig.~\ref{fig:dndz}).

The potential of using the \sze\ as a tool to help determine
the cosmological parameters is now being realized nearly 
three decades after it was first proposed. We expect improvements
to move forward at a rapid pace with sensitive \sze\ surveys of the
the high redshift universe starting in the next few years.

\bigskip
We thank the OVRO and BIMA observatories for their
crucial contributions to the \sze\ observations. JC thanks 
organizers of the symposium for 
a remarkably informative and enjoyable symposium. JC acknowledges
support from a NSF-YI grant and the David and Lucile Packard
Foundation.

\bibliographystyle{aas}
\bibliography{carlstrom}

\begin{thebibliography}{}

\bibitem[Bahcall 1999]{bahcall99}
Bahcall, N. 1999.
\newblock Cosmology with Clusters of Galaxies.
\newblock In Bergstrom, L., Carlson, P., and Fransson, C., editors, {\em Nobel
  Symposium on Particle Physics and the Universe}, Sweden. Physics Scripta --
  astro-ph/9901076.

\bibitem[Barbosa {\em et~al.} 1996]{barbosa96}
Barbosa, D., Bartlett, J., Blanchard, A., and Oukbir, J. 1996, \aap, {\bf 314},
  13.

\bibitem[Birkinshaw 1991]{birkinshaw91}
Birkinshaw, M. 1991, {\it Physical Cosmology}, ed. J. Tran Thanh Yan, Editions
  Frontiers, page 177.

\bibitem[Birkinshaw 1999]{birkinshaw99}
Birkinshaw, M. 1999, Physics Reports, {\bf 310}, 97.

\bibitem[{Burles} and {Tytler} 1998]{burles98a}
{Burles}, S. and {Tytler}, D. 1998, \apj, {\bf 507}, 732.

\bibitem[Carlstrom {\em et~al.} 1998]{carlstrom98}
Carlstrom, J.~E., Grego, L., Holzapfel, W.~L., and Joy, M. 1998, {Eighteenth
  Texas Symposium on Relativistic Astrophysics and Cosmology}, ed A. Olinto, J.
  Frieman, and D. Schramm, World Scientific, page 261.

\bibitem[Carlstrom, Joy, and Grego 1996]{carlstrom96}
Carlstrom, J.~E., Joy, M., and Grego, L.~E. 1996, \apjl, {\bf 456}, L75.

\bibitem[{Colafrancesco} {\em et~al.} 1997]{colafrancesco97}
{Colafrancesco}, S., {Mazzotta}, P., {Rephaeli}, Y., and {Vittorio}, N. 1997,
  \apj, {\bf 479}, 1.

\bibitem[Cooray {\em et~al.} 1998]{cooray98}
Cooray, A.~R., Grego, L., Holzapfel, W.~L., Joy, M., and Carlstrom, J.~E. 1998,
  \aj, {\bf 115}, 1388.

\bibitem[David, Jones, and Forman 1995]{david95}
David, L., Jones, C., and Forman, W. 1995, \apj, {\bf 445}, 578.

\bibitem[Evrard 1997]{evrard97}
Evrard, A. 1997, \mnras, {\bf 292}, 289.

\bibitem[{Grainge} {\em et~al.} 1996]{grainge96}
{Grainge}, K., {Jones}, M., {Pooley}, G., {Saunders}, R., {Baker}, J.,
  {Haynes}, T., and {Edge}, A. 1996, \mnras, {\bf 278}, L17.

\bibitem[{Grainge} {\em et~al.} 1993]{grainge93}
{Grainge}, K., {Jones}, M., {Pooley}, G., {Saunders}, R., and {Edge}, A. 1993,
  \mnras, {\bf 265}, L57.

\bibitem[Grego 1999]{grego99a}
Grego, L. 1999.
\newblock {\em Galaxy Cluster Gas Fractions from Interferometric Measurements
  of the Sunyaev-Zel'dovich Effect}.
\newblock PhD thesis, Caltech.

\bibitem[Grego {\em et~al.} 1999a]{grego99c}
Grego, L., Carlstrom, J.~E., Joy, M.~K., Reese, E.~D., Holder, G.~P., Patel,
  S., Cooray, A.~R., and Holzapfel, W.~L. 1999a, ApJ -- submitted.

\bibitem[Grego {\em et~al.} 1999b]{grego99b}
Grego, L., Carlstrom, J.~E., Joy, M.~K., Reese, E.~D., Holder, G.~P., Patel,
  S., Cooray, A.~R., and Holzapfel, W.~L. 1999b, ApJ -- submitted.

\bibitem[{Herbig} {\em et~al.} 1995]{herbig95}
{Herbig}, T., {Lawrence}, C.~R., {Readhead}, A. C.~S., and {Gulkis}, S. 1995,
  \apjl, {\bf 449}, L5.

\bibitem[Holder and Carlstrom 1999]{holder99a}
Holder, G. and Carlstrom, J. 1999.
\newblock The Sunyaev-Zeldovich Effect as Microwave Foreground and Probe of
  Cosmology.
\newblock In de~Oliveira-Costa, A. and Tegmark, M., editors, {\em Microwave
  Foregrounds}, San Francisco. ASP-- astro-ph/9904220.

\bibitem[Holder, Carlstrom, and Mohr 1999]{holder99b}
Holder, G., Carlstrom, J., and Mohr, J. 1999, ApJ -- in preparation.

\bibitem[Holzapfel {\em et~al.} 1997a]{holzapfel97b}
Holzapfel, W.~L., Ade, P.~A.~R., Church, S.~E., Mauskopf, P.~D., Rephaeli, Y.,
  Wilbanks, T.~M., and Lange, A.~E. 1997a, \apj, {\bf 481}, 35.

\bibitem[Holzapfel {\em et~al.} 1997b]{holzapfel97a}
Holzapfel, W.~L. {\em et~al.} 1997b, \apj, {\bf 480}, 449.

\bibitem[{Jones} {\em et~al.} 1993]{jones93}
{Jones}, M. {\em et~al.} 1993, Nature, {\bf 365}, 320.

\bibitem[{Lamarre} {\em et~al.} 1998]{lamarre98}
{Lamarre}, J.~M. {\em et~al.} 1998, \apjl, {\bf 507}, L5.

\bibitem[Mohr, Mathiesen, and Evrard 1999]{mohr99}
Mohr, J., Mathiesen, B., and Evrard, A. 1999, \apj, {\bf 517}, --
  astro--ph/9901281.

\bibitem[Myers {\em et~al.} 1997]{myers97}
Myers, S.~T., Baker, J.~E., Readhead, S., A.~C., Leitch, E.~M., and Herbig, T.
  1997, \apj, {\bf 485}, 1.

\bibitem[Perlmutter {\em et~al.} 1999]{perlmutter99}
Perlmutter, S. {\em et~al.} 1999, \apj, pages accepted, see astro--ph/9812133.

\bibitem[Reese {\em et~al.} 1999]{reese99}
Reese, E.~D. {\em et~al.} 1999, ApJ -- in preparation.

\bibitem[Rephaeli 1995]{rephaeli95}
Rephaeli, Y. 1995, \araa, {\bf 33}, 541.

\bibitem[Riess {\em et~al.} 1998]{riess98}
Riess, A.~G. {\em et~al.} 1998, \aj, {\bf 116}, 1009.

\bibitem[Sunyaev and Zel'dovich 1970]{sunyaev70}
Sunyaev, R. and Zel'dovich, Y. 1970, Comments Astrophys. Space Phys., {\bf 2},
  66.

\bibitem[Sunyaev and Zel'dovich 1972]{sunyaev72}
Sunyaev, R. and Zel'dovich, Y. 1972, Comments Astrophys. Space Phys., {\bf 4},
  173.

\bibitem[Sunyaev and Zel'dovich 1980]{sunyaev80}
Sunyaev, R. and Zel'dovich, Y. 1980, ARAA, {\bf 18}, 537.

\bibitem[Vikhlinin {\em et~al.} 1998]{vikhlinin98}
Vikhlinin, A., McNamara, B., Forman, W., Jones, C., Quintana, H., and
  Hornstrup, A. 1998, \apj, {\bf 502}, 558.

\bibitem[{Wilbanks} {\em et~al.} 1994]{wilbanks94}
{Wilbanks}, T.~M., {Ade}, P. A.~R., {Fischer}, M.~L., {Holzapfel}, W.~L., and
  {Lange}, A.~E. 1994, \apjl, {\bf 427}, L75.

\end{thebibliography}

\end{document}